# Intermittency of Internal Wave Shear and Turbulence Dissipation

Carl H. Gibson
Departments of Applied Mechanics and Engineering Sciences
and Scripps Institution of Oceanography, University of California at San Diego
La Jolla, California, 92093-0411, USA, cgibson@ucsd.edu

Abstract

It is crucial to understand the extreme intermittency of ocean and lake turbulence and turbulent mixing in order to estimate vertical fluxes of momentum, heat and mass by Osborn-Cox flux-dissipation methods. Vast undersampling errors occur by this method when intermittency is not taken into account, Gibson (1990a, 1991ab). Turbulence dissipation and internal wave shear in the ocean are closely coupled. Often, oceanic turbulence is assumed to be caused by breaking internal waves. However, the extreme intermittency observed by Gregg et al. (1993), with intermittency factor $I_{S^4}$ 6 for 10 m internal wave shears S, strongly suggests that wave breaking is not the cause of turbulence but an effect. The usual assumption about the wave-turbulence cause-effect relationship should be reversed. Wave motions alone reduce intermittency. Oceanic turbulence increases intermittency as the result of a self-similar nonlinear cascade covering a wide range of scales, mostly horizontal, but no such nonlinear cascade exists for internal waves. Extremely large I and I measured for oceanic turbulence are in the range 3-7, Baker and Gibson (1987). These values are consistent with the third universal similarity hypothesis for turbulence of Kolmogorov (1962) and a length scale range over 3-7 decades from viscous or diffusive to buoyancy or Coriolis force domination, where the measured universal intermittency constant $\mu$ = 0.44, Gibson (1991a), is a result of singularities in multifractal turbulence dissipation networks and their degeneration, Bershadskii and Gibson (1994). The extreme intermittency of small scale internal wave shears in the ocean is a fossil turbulence remnant of the extreme intermittency of ocean turbulence, and the waves themselves may be fossil turbulence. Evidence that composite shear spectra and schematic temperature gradient spectra in the ocean reflect fossil turbulence effects, Gibson (1986), is provided by the decaying tidal sill temperature spectra of Rodrigues-Sero and Hendershott (1997).

## Introduction

Davis (1994) has reviewed the flux-dissipation method, referred to as the Osborn-Cox method when applied to the ocean, to estimate vertical diffusivites $K_V$ from the expression $K_V$ DC, where D is the molecular diffusivity of a conserved property like species concentration or temperature T (internal energy) and C is the Cox number C = ( )$^2$/ $^2$, and points out that large discrepancies exist between such microstructure values and much larger values found by matching large-scale observations to models or budgets; e. g., Munk (1966). Several possible sources of small errors in $K_V$ are considered, but not intermittency in space-time versus control volumes as emphasized by Gibson (1990a) in another review of the Osborn-Cox method.  Davis (1996) considers some of the possible effects of intermittency on $K_V$ estimates, but criticizes the suggestion of Baker and Gibson (1987) that since most microstructure data sets for , and C in identifiable ocean layers have sample distribution functions imperceptibly different from intermittent lognormal distributions, that maximum likelihood estimators of mean values based on lognormality should provide the best available estimators for $K_V$.

Davis (1996) suggests | | is more likely to be lognormal than  and that if | | is lognormal then  and averages of  over various length scales cannot be lognormal, contrary to the proposal by Gibson (1981) that  and averages of  should be lognormal for oceanic mixing over a wide range of length scales.  The Gibson (1981) analysis simply extends to turbulent mixing of conserved scalar fluid properties the very robust Gurvich and Yaglom (1967) breakdown coefficient analysis that provides the physical basis for Kolmogorov's third universal similarity hypothesis of intermittent lognormality of turbulence dissipation rates.  This suggestion of Davis (1996) has no physical basis and is contrary to observations of  averaged over scales $L_{ave.}$ from 1 to 1024 meters in the seasonal thermocline off the coast of California, Gibson (1991a), that were precisely lognormal with I  0.44 ln ($L_O/L_{ave.}$) for all $L_{ave.}$.

The Davis (1996) suggestion that fitting  observations to lognormal distributions can give quite erroneous results is inconsistent with Gibson (1991b) estimates of $K_V$ values from the 150 m average C values reported by Gregg (1977).  These agree well with lognormality for various ocean layers sampled, and confirm the Munk (1966) estimates of $K_V$ for vertical heat flux in the deep main thermocline, contrary to the Gregg (1989) correlation that underestimates the deep heat flux by a factor of about thirty.  No statistically significant difference exists between $K_V$ values inferred by the Munk (1966) abyssal recipe and the flux-dissipation methods (Osborn-Cox) based on the Gregg (1977) data, as shown by Gibson (1991a) taking the lognormal intermittency of C into account.

Some physical processes of nature are intermittent and some are not, where the degree of intermittency of a random variable X is measured by statistical parameters showing departures of its distribution function from that of a normal distribution; for example, skewness, kurtosis, and the intermittency factor $I_X$, where

$$I_X \quad ^2_{lnX}, \; X \; lognormal, \qquad [1]$$

is defined by Baker and Gibson (1987). What is the difference? Examples of non-intermittent processes in the ocean include the temperature, pressure, salinity, most chemical species distributions, and surface and internal wave displacements. Examples of intermittent oceanic processes include the viscous dissipation rate  of turbulence, the thermal diffusive dissipation rate of temperature variance , the Cox number C = ( )$^2$/ $^2$ for temperature, the shear rate S of small scale internal waves in the ocean, the rate of biological productivity, and the vertical fluxes of heat, mass of chemical species, and momentum through the thermocline, Gibson (1990a). Why are these various examples different with respect to their intermittency?

Part of the difference has to do with the differences in intermittency of the sources of the various physical parameters. Temperature differences in the ocean are primarily determined by smooth large-scale variations in radiation transport on scales of the planet, and many of the non-intermittent chemical species distributions, like oxygen, also have uniform very large-scale sources and sinks. Chemical species with intermittent sources in space and time are more intermittent than those that have non-intermittent sources and sinks. As soon as the intermittent source is removed, the intermittency of such species concentrations begin to disappear as their maxima are damped and dispersed by diffusion and convection. Conservation laws for such random variables are linear in the variable and will tend to have normal probability distribution functions unless forced away from Gaussianity by non-Gaussian, intermittent, source terms. For example, tritium concentration in the ocean today is less intermittent than it was in the days of intermittent nuclear weapons testing. Biological growth rates are extremely intermittent because they depend on a coincidence of several random variables (such as nutrient concentrations, sunlight, turbulence, and even turbulence intermittency, Gibson and Thomas (1995)) having their proper value ranges simultaneously. Deviations from intermittent lognormality of oceanic dissipation rates occur when data sets include segments from different layers with different intermittency parameters, Yamazaki and Lueck (1990).

According to the intermittency hypothesis of Kolmogorov (1962), random variables of turbulence in the ocean such as are highly intermittent because the equations of turbulent momentum and vorticity are highly nonlinear and the range of scales of the turbulence cascade is enormous. Universal intermittency parameters of turbulence have been investigated in terms of singularities on dissipation networks, or caustics, and their degeneration, by Bershadskii and Gibson (1994), using multifractal asymptotics. The hallmark of intermittent random variables is a self-similar nonlinear cascade over a wide range of scales. Classic examples are turbulence, turbulent mixing, personal income, concentrations of precious metals, and the density of astrophysical fluids undergoing self gravitational condensation. Distributions of these random variables are close to intermittent lognormals. Each is subject to severe undersampling errors.

The intermittency parameter kurtosis K is defined as the fourth moment of X about its mean divided by its second moment, where

$$K = \exp(I_X) \qquad [2]$$

for a lognormal random variable X. For a normal random variable, K = 3. Therefore, from [2], $I_X$ is 1.1 for normal random variables. Most data sets from particular oceanic layers show that the probability distributions of the viscous and temperature dissipation rates and are indistinguishable from intermittent lognormal distributions, Baker and Gibson (1987), and have intermittent I and I values in the range 3-7, much larger than 1.1. Such large departures from Gaussian behavior can produce severe problems for any experimental attempt to sample mean values of and because the mean to mode ratio G increases exponentially with $I_X$

$$G = \text{Mean/Mode} = \exp(3I_X/2). \qquad [3]$$

The ratio G is the expected undersampling error for intermittent random variables using sparse samples, and might be termed the Gurvich number. G equals 90 for $I_X = 3$, and $4 \times 10^4$ for $I_X = 7$. Thus, the expected undersampling errors for single samples of oceanic $\varepsilon$ or $\chi$ are underestimates by two to four orders of magnitude. The first claim that the Munk (1966) abyssal recipe was incorrect was that of Osborn and Cox (1972), based on a single 25 m long microstructure record. Confirmation of this erroneous claim was provided by Gregg, Cox and Hacker (1973) with similar measurements of small C values (1.3, 2.2, 1.3, 2.0, 1.7 and 1.4) in the North Pacific Central Gyre, and 286 in the San Diego Trough neglecting G corrections.

The possibility of large undersampling errors demands extreme care in estimating mean dissipation rates from sparse data sets. Large numbers of samples are needed, but the samples must be independent and representative of the full range of space and time scales in the space-time control volume of interest. As mentioned above, Gibson (1991b) examines relatively independent 150 m average Cox numbers C, where C can be interpreted as the ratio of the average turbulent to molecular vertical diffusivities, measured by Gregg (1977) in three layers of the upper ocean of the mid-Pacific for three seasons of three years, and finds C to be an intermittent lognormal with $I_C$ values in the range 3-6 for all layers. No statistically significant difference was found between the maximum likelihood estimator of the mean $\overline{C}$ and the value $\overline{C} \approx 700$ predicted by Munk (1966) for deep vertical diffusivity in the main thermocline.

Mode values of C are smaller than $\overline{C} \approx 700$ by factors of 1-2 orders of magnitude according to numerous investigators using microstructure dropsondes, leading to a "dark mixing" paradox (a term invented by Tom Dillon) analogous to the "dark matter" paradox of cosmology. No "dark mixing" paradox exists for $\overline{C}$ in the ocean if its extreme intermittency is taken into account. The Gregg (1995) conclusions that $\varepsilon$, $\chi$, and $\overline{C}$ are less near the equator than at mid-latitudes may actually reflect maximum undersampling errors (maximum Gurvich numbers) in dropsonde microstructure methodology near the equator where the intermittency factors for these quantities are maximum, due to the minimum in Coriolis forces and the huge range of horizontal turbulence scales up to hundreds of kilometers, Gibson (1983).

Any hydrophysical field of the ocean that is generated by an intermittent source will itself be intermittent, at least initially. Random variables like dissipation rates $\varepsilon$ and $\chi$ are intrinsically intermittent because turbulence is intermittent, with $I_\varepsilon$ and $I_\chi$ values increasing $\propto \ln L_O$. Random variables like temperature, internal wave shear, and chemical species concentration can disperse widely, and will be intermittent only as long as they are forced to be so by intermittent sources. Clearly if X is an intermittent random variable, then a random variable $Y \propto X$ will have the same intermittency parameters as X because Y has the same probability density function as X. Proportionality constants c cancel for the variance of logarithms, so that $I_X = I_x$ if $X = cx$. Other random variables $Z = f(X)$ will be more or less intermittent than X depending on the function $f(X)$. Intermittency factors $I_\chi$ and $I_\varepsilon$ can easily be measured because $I_\chi = I_{(\partial T/\partial x)^2}$ and $I_\varepsilon = I_{(\partial u/\partial x)^2}$ since $\chi \propto (\partial T/\partial x)^2$ and $\varepsilon \propto (\partial u/\partial x)^2$.

Measured intermittency factors $I_\chi$ not only reflect measured large intermittency factors $I_\varepsilon$ but have the same values, Gibson (1991a). A model predicting this equality is given by Gibson (1981) based on the Gurvich and Yaglom (1967) model for the Kolmogorov (1962) intermittency hypothesis. Small scale internal wave shear S, $S^2$, $S^4$, ... , should have small $I_{S^4}$ if the shear is dominated by linear wave mechanics and large $I_{S^4}$ if small scale internal waves are caused by turbulence. Intermittency factors are good indicators of whether the processes that produce a particular random variable are linear or nonlinear. If $I_{S^4}$ is large, it does not follow that turbulence is caused by breaking waves. Instead, it follows that the waves are caused by turbulence because the intermittency of the wave shear reflects the intermittency of the turbulence. If the turbulence were caused by waves $I_\varepsilon$ values would be near 1.1, not 3-7.

Gregg et al. (1993) show that the internal wave shear S on scales of about 10 m is quite intermittent, and that $S^4$ is lognormal with $I_{S^4} \approx 6$. Because $\varepsilon \sim S^4$ according to Gregg (1989), Gregg et al. (1993) suggest the intermittency of $\varepsilon$ may be explained as due to the intermittency of $S^4$ assuming turbulence is caused by breaking internal waves, but this conclusion is precisely reversed. Internal wave shears cannot be intermittent unless the waves are forced by nonlinear processes, and the most likely nonlinear forcing process for the ocean interior is turbulence. Waves break when they are forced to do so by the surface, beaches, or bottom topography (or by the turbulence triggered by these entities). Internal wave shears on 10 m scales are fossils of the rare turbulence events that intermittently reach such large vertical scales, Gibson (1986).

The Gregg et al. (1993) observation that $I_{S^4} \approx 6$ can be compared to a measured intermittency factor $I_\varepsilon \approx 5$, Washburn and Gibson (1984). $S^4 \sim \varepsilon^{4/3} L^{-8/3}$ from Kolmogorov's second hypothesis, since $V \sim (\varepsilon L)^{1/3}$ and $S \sim (\Delta V)/L$. Therefore,

$$I_{S^4} \approx (4/3)^2 I_\varepsilon . \qquad [4]$$

from the definition of $I_X$. According to Kolmogorov's third hypothesis, $I_\varepsilon \approx 0.44 \ln (L_O/cr)$, where $L_O$ is the energy, or Obukhov, scale of the turbulent cascade, r is the averaging scale for $\varepsilon$, and c is a constant of order ten. Taking cr = 0.3 m from Washburn and Gibson (1984) gives $L_O \approx 26$ km from $I_\varepsilon \approx 5$, and $I_\varepsilon$ only 2.4 for cr = 100 m from [4], compared to $I_\varepsilon = 6/(4/3)^2 = 3.5$ from Gregg et al. (1993). This suggests a larger $L_O$ value for Gregg et al. (1993).

Active turbulence in the ocean interior has never been observed simultaneously with internal waves with the extremely large shear rates S that must exist to produce the extreme intermittency factors $I_{S^4} \approx 6$ reported by Gregg et al. (1993). Fossil turbulence remnants of previous powerful mixing events, with vertical overturns up to 8 m, have been identified by Gibson (1982) from the Gregg (1977) profiles with largest C values, but S values were not measured. Wave spectra far downstream of powerful turbulence on a tidal sill are given below.

Temperature and density overturns on 20 m scales in the strong shear layer of the equatorial undercurrent were detected by Hebert et al. (1992) in a large patch of partially fossilized turbulence precisely in the center of a packet of internal waves with maximum amplitude of 20 m. Their claim that the strong simultaneous (but partially fossilized) turbulence observed was caused by the breaking of the 20 m amplitude internal wave packet they measured is questioned by Gibson (1991b) on the grounds that turbulence overturns caused by breaking waves should be on scales smaller than the amplitude of the wave and not identical to it. Identical maximum amplitudes for overturning scales and internal waves is a signature of internal waves caused by turbulence, Gibson (1980, 1986). Internal waves in the ocean interior typically have amplitudes of tens of meters, but turbulence overturn scales are ten meters or less. Recent measurements of Rodrigues-Sero and Hendershott (1997) showing evidence of saturated internal waves and persistent 60 m vertical amplitude fossil turbulence temperature overturns in the Sea of Cortez are described below. These internal waves are known to have been produced by powerful turbulence formed by tides flowing over sills, with overturn scales identical to the amplitude of the internal waves produced.

## Causes of Intermittency in Physical Systems

The most important source of intermittency in physical systems is nonlinearity of the conservation equations describing the random variable. For the case of turbulence, turbulence dissipation rate , small scale internal waves, $S^4$ etc., the random variable in the ocean determining these random variables is the vorticity , that is described by the equation

$$\frac{\partial \omega}{\partial t} + (\mathbf{v} \cdot \nabla) \omega = \omega \cdot \mathbf{e} + 2\Omega \cdot \mathbf{e} + \frac{\nabla \rho \times \nabla p}{\rho^2} - \frac{\nabla \times \nu \nabla^2 \mathbf{v}}{\nu} + \nu \nabla^2 \omega \quad [5]$$

where **e** is the rate of strain tensor with components ($\partial v_i/\partial x_i + \partial v_j/\partial x_i$)/2, $\Omega$ is the rotation vector of the coordinate system, $\rho$ is density, p is pressure, and $\nu$ is the kinematic viscosity. The left hand side of [5] is the rate of change of the vorticity of a fluid particle.

The vorticity becomes intermittent because of the nonlinearity of the first term on the right hand side of [5], representing the rate of increase due to vortex stretching. The term is nonlinear in velocity gradients, and causes rapid increases in vorticity magnitude because regions with large vorticity acquire large strain rates in the stretching direction when the vorticity aligns with the stretching axis of the strain rate tensor as it is amplified by it. The second (Coriolis) term on the right is generally smaller than the first. The third term shows that vorticity is produced on tilted density interfaces such as occur in small scale internal waves. However, the vorticity produced tends to disappear and reverse for internal waves with small amplitude that do not form turbulence.

When the density gradient tilt is steady, as on a front, the shear can build up until a turbulence patch forms. Such persistent density gradient tilts are characteristic of fronts formed at boundaries of horizontal turbulence eddies stirring horizontal density gradients, as shown in Figure 1. The relationship of equation [4] shown in Fig. 1 between $I_{S^4}$ and $I_\varepsilon$ assumes equal averaging scales are used for both the shear S and the viscous dissipation rate . From Kolmogorov's third hypothesis $I_\varepsilon = 0.44 \ln (L_O/cr)$ and the relation $I_{S^4} \approx (4/3)^2 I_\varepsilon$ it is clear that such large values of shear intermittency factors as $I_{S^4} \approx 6$ on scales of 10 m require such a large range of scales in the turbulence cascade, to tens of kilometers, that most of the turbulence energy must be in the horizontal rather than in the vertical.

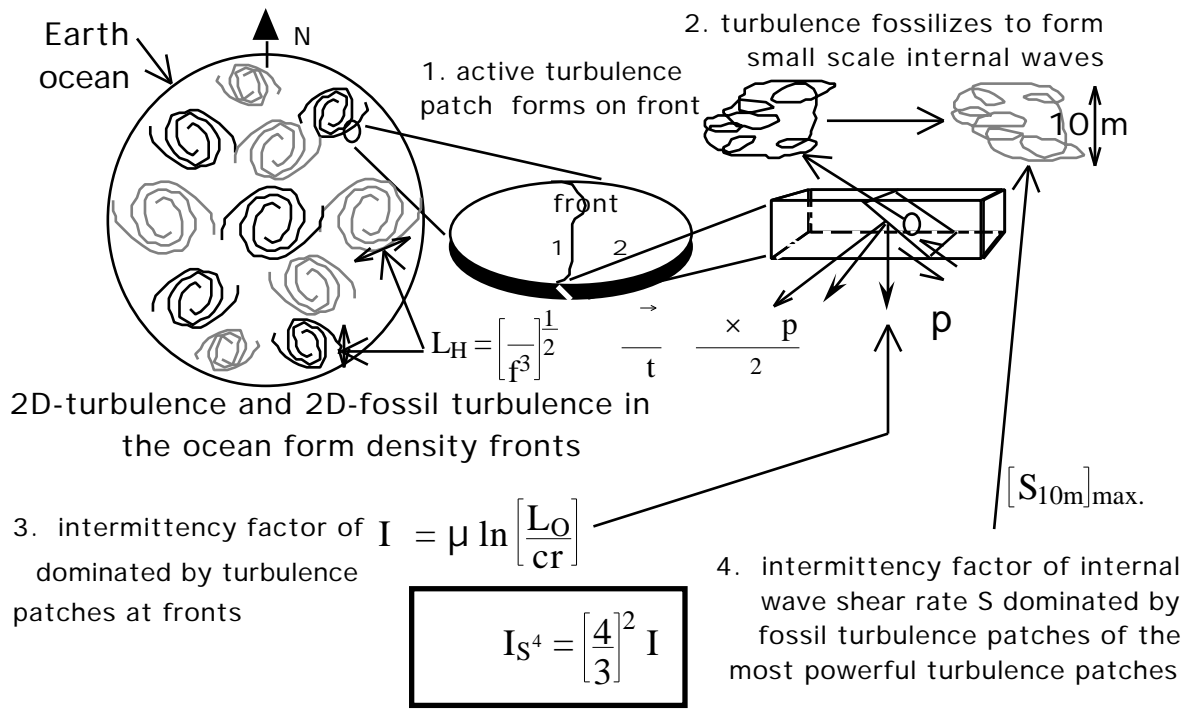

**Figure 1.** Four step model showing how the intermittency of ε causes the intermittency of internal wave shear rates $S^4$ if the waves are fossil vorticity turbulence.

As shown in Fig. 1, horizontal scales of turbulence in the ocean are limited either by the time of application of the source as turbulence cascades from small scales to large, or by the Hopfinger scale $L_H = (\varepsilon/f^3)^{1/2}$, where f is the Coriolis parameter $2\Omega \sin \phi$ and $\phi$ is latitude. Horizontal eddy sizes up to hundreds of kilometers are observed at equatorial latitudes, compared to tens of kilometers near the poles, and intermittency factors estimated by Baker and Gibson (1987) show a similar correlation. It will be interesting to see whether values of the shear intermittency factor $I_{S^4}$ are also correlated with latitude. This would be further evidence that the dominant small scale internal wave shear is caused by the dominant turbulence events (those that control the vertical fluxes), rather than such turbulence being caused by internal wave breaking.

As mentioned previously, Gurvich and Yaglom (1967) provide a clear physical model for the intermittent, lognormal, distribution function for the dissipation rate ε of very high Reynolds number turbulence postulated by Kolmogorov (1962) in his third universal similarity hypothesis for turbulence. This model and Kolmogorov's hypothesis have been extended to passive scalar fields like temperature mixed by turbulence, Gibson (1981). Both models note that the mechanisms of turbulence and turbulent mixing are identical for every stage of the nonlinear cascades from small scales to large, and that every stage is independent of much larger or much smaller stages. If ε is averaged over a region of size r it should be independent of ε averaged over a larger region of size nr for some value of n>1. The ratio of $\varepsilon_r$ to $\varepsilon_{nr}$ and the ratio $\varepsilon_{nr}$ to $\varepsilon_{n^2r}$ should be independent and identically distributed. Therefore, the ratio $\varepsilon_r$ to $\varepsilon_{Lo}$ can be written as the product of a large number of independent, identically distributed random variables with the form $\varepsilon_{n^k r} / \varepsilon_{n^{k+1} r}$ for k an integer >1 if there is a very large range of scales between r and the energy, or Obukhov, scale Lo, Gibson (1990b).

Taking the logarithm of the product and applying the central limit theorem shows that $\varepsilon_r$ must be lognormally distributed with intermittency factor increasing logarithmically with the Reynolds number, as assumed by the Kolmogorov (1962) third universal similarity hypothesis that the viscous dissipation rate averaged over scale r should be lognormal, with variance given by

$$I = \sigma^2_{\ln \varepsilon_r} = \mu \ln(Lo/cr) \qquad [6]$$

where µ is a universal constant and c is a constant depending on the geometry of the energy scales. Measurements in the atmospheric boundary layer over the open ocean gave values of µ = 0.5± 0.1, Gibson, Stegen and McConnell (1970) from slopes of spectra of the dissipation E ~ $k^{-1+\mu}$, and µ = 0.47±0.03 from direct tests of [6] using different averaging intervals r, Gibson and Masiello (1971). Measurements of temperature dissipation rate $\chi$ in the seasonal thermocline of the California upwelling region in Monterey Bay by oceanographers of the former Soviet Union on the AKADEMIK KURCHATOV, Iosif Lozovatsky Chief Scientist, reduced by Mark Baker, gave values for the temperature equivalent of equation [6] of µ = 0.44±0.01, with very precise agreement of $\chi$ averaged over length scales at each octave from a meter to a kilometer with lognormal distributions, Gibson (1991a). Falgarone and Phillips (1990) inferred µ = 0.4-0.5 from measurements of line broadening in a dense molecular cloud of the galaxy attributed to intermittent turbulence, with $\chi$ values comparable to those in the upper ocean, Gibson (1991a). Structure functions were computed for separation lengths of $10^{15}$ to $10^{19}$ meters within this large non-star-forming gas cloud.

Estimation of the Kolmogorov Intermittency Index µ for the Ocean

Several authors have criticized the Kolmogorov (1962) hypothesis of lognormal intermittency for $\varepsilon$ and the Gurvich and Yaglom (1967) physical justification; for example, Meneveau and Sreenivasan (1993). Exponential distributions have been predicted rather than lognormal, and claims of multifractal behavior of turbulence with no universal intermittency coefficient µ.

Bershadskii and Gibson (1994) show that µ = 0.5 is a limiting value to be expected at very high Reynolds numbers Re, and that the lognormality of $\varepsilon$ at such high Re follows from multifractal asymptotics. Turbulence dissipation at low Re is concentrated on caustic networks that appear due to vortex sheet instability in three dimensional space, leading to an effective fractal dimension D of 5/3 of the network backbone without caustic singularities and a turbulence intermittency exponent µ = 1/6. As Re increases stable singularities form on the caustics, so that D decreases and µ increases. Four stable types of caustic singularities are identified: cusp ridge, swallowtails, pyramid, and purse points, Arnold (1990, 1986). For sheet-like caustics, cusp ridge (ordinary) singularities have $D = 3/2$ and $\mu = 1/4$; swallowtails $D = 1.4$ and $\mu = 0.3$; and pyramid and purse points, $D = 4/3$ and $\mu = 1/3$. Further increases of Re cause degeneration of the caustic networks into smooth vortex filaments with $D = 1$ and $\mu = 1/2$. The conclusion is that low Re numerical simulations of turbulence intermittency are likely to indicate $\mu$ values of 0.2, whereas field measurements are likely to indicate $\mu$ values near 0.5, as observed for atmospheric, oceanic and astrophysical turbulence with very long cascade ranges.

Figure 2 shows the Obukhov, or energy, scale $L_O$ calculated from [6] as a function of I for $\mu = 0.44$ and 0.2 for cr = 0.3 m. The energy scale $L_O$ indicated for a given I is very sensitive to the value selected for $\mu$. Gurvich numbers G range from 20 to $3.3 \times 10^6$, from [3].

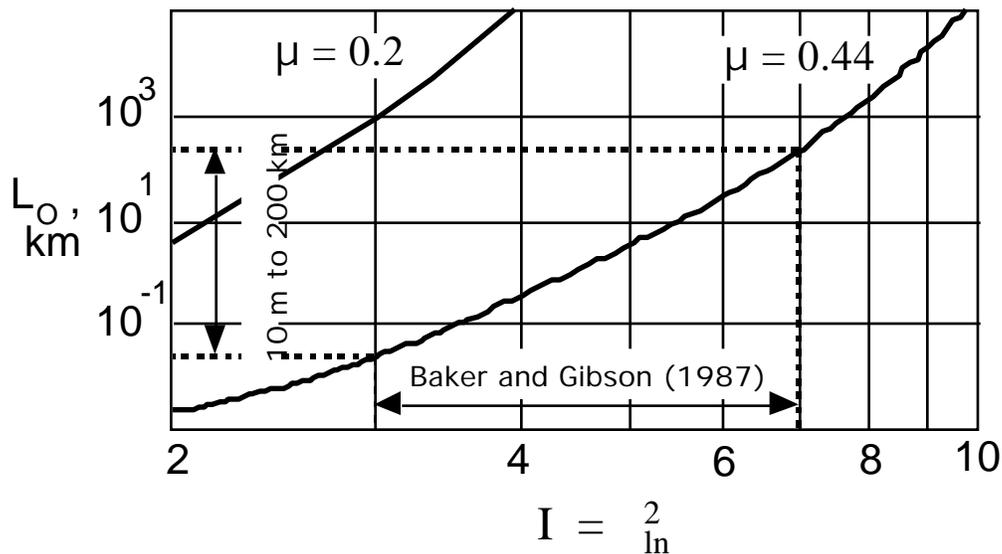

**Figure 2.** Energy scale $L_O$ versus intermittency factor I for $\mu = 0.44$ and 0.2 with cr = 30 cm, from [6]. A reasonable range of $L_O$ is found for $\mu = 0.44$, but not for $\mu = 0.2$.

Smaller values of $\mu = 0.25$ and less have been inferred by other authors, Sreenivasan and Kailasnath (1993), usually from data with much lower Reynolds number, or from statistical estimates based on length scales outside the universal range (which is on much smaller scales for than it is for the universal range of velocity u, by a factor of about 7). Wang et al. (1996) find $\mu$ values in the range 0.20 to 0.28 using a massively parallel connection machine with Taylor microscale Reynolds number of about 200. The dashed lines show the range of energy scales implied by the Baker and Gibson (1987) intermittency factor estimates from ocean measurements are in the range 10 m to 200 km for $\mu = 0.44$, Gibson (1991a), which is reasonable. For $\mu = 0.2$ the range is from 1000 km to $4 \times 10^{11}$ km, which is not reasonable.

Since some in the oceanographic community may not have accepted the fact that intermittency has crucial effects on the proper sampling of oceanic turbulence and turbulent mixing dissipation rates needed to apply the Osborn-Cox method, it might be instructive to review other self-similar nonlinear cascade processes that lead to intermittent random variables (and undersamplng errors) with different efficiencies. For example, the more money you have the more rapidly you can get more, especially if you have lots. Personal income rates for the not-so-rich (the lower 97%) is lognormal, but with $I_\$ = \sigma^2_{\ln \$}$ of only 0.31 compared to $I_\$ = 4.6$ for the super-rich (the upper 3%), Montroll and Shlesinger (1982). Tax collectors in Sultanates will clearly fail if they neglect to tax the Sultan and his richest friends.

Self-gravitational condensation of density in the universe is an extreme example, with variations of density covering a range of more than 50 decades, from intergalactic supervoid densities of $10^{-35}$ kg m$^{-3}$ to neutron star densities of $10^{15}$ kg m$^{-3}$. Hubble (1934) showed that the density of galaxies is lognormal by counting over 44,000 "extra-galactic nebulae" on 1283 plates taken of three quarters of the sky north of -30 degrees declination. Galaxy densities measured by Hubble were smoothed over 639 fields and had intermittency factor $I_\rho$ of only about 1.0, although intermittency factors $I_\rho$ for modern red-shift sorted catalogs and cold-dark-matter calculations are much larger, up to 9, Columbi (1994), with an intermittency exponent µ about 2.5 compared to about 0.5 for turbulence. Estimating the average density of the universe from a small number of random samples is even more futile than estimating $\varepsilon$ or $\chi$ in the ocean by this method, except that the Gurvich number for $\rho$ is likely to be $10^9$, underestimating the true value of $10^{-26}$ kg m$^{-3}$ by a billion, rather than too small by only factors of $10^1$ to $10^4$ typical of oceanic undersampling errors for average C, $\varepsilon$ and $\chi$, Gibson (1996a). Estimation of the density of "massive compact halo objects" (MACHOs) in the galaxy halo by microlensing of nearby star fields gives smaller values than densities estimated by quasar microlensing which averages over larger halo volumes, when the extreme intermittency of MACHO density is not taken into account, Gibson (1996a). Part of the dark matter paradox is undersampling error.

## Intermittency of Internal Waves and Turbulence in the Ocean

"*Because breaking internal waves produces most of the turbulence in the thermocline, the statistics of $\varepsilon$, the rate of turbulent dissipation, cannot be understood apart from the statistics of internal wave shear*" is the leading sentence of the Abstract of Gregg, Seim and Percival (1993). No proof is offered for the assumption that most of the turbulence in the thermocline is produced by breaking internal waves, but this is a crucial step in their goal of comparing the statistics of small scale internal wave shear and the statistics of $\varepsilon$. Later in the Abstract the authors state, "*It is hypothesized that the approximate lognormality of bulk ensembles of $\varepsilon$ results from generation of turbulence in proportion to $S^4$*", where S is the internal wave shear measured over a 10 m vertical interval. This hypothesis depends on the previous assumption that turbulence is caused by waves, and is equally questionable. Certainly in some circumstances the reverse relationship will occur, where internal waves are caused by turbulence. Why should internal wave shear be intermittent without turbulence?

Internal waves, like surface waves, are generally governed by linear equations at the largest scales. Vertical displacements can be represented as the sum of a large number of independent random variables, and have normal probability distributions as expected from the central limit theorem. Wave energy disperses over larger and larger volumes and surface areas, reducing any nonlinear tendencies. Only at the smallest scales does nonlinearity develop. At small scales internal waves are described by the same equations as turbulence, and merge with turbulence at that scale. If large scale internal waves are forced to large amplitudes they develop thinner and thinner density steps with stronger and stronger density gradients that inhibit the formation of turbulence, Phillips (1969). The shear for a given patch of water subjected to internal wave motions increases and decreases reversibly. Concentrations of energy in a wave field are propagated to larger fluid volumes by well known linear mechanisms.

Clearly turbulence can generate internal waves in a stratified fluid. Mechanisms include the formation of fossil vorticity turbulence on scales smaller than those of the most active turbulence, Gibson (1980, 1986), and on larger scales external to the turbulent fluid, Townsend (1965, 1966), Carruthers and Hunt (1986), Uittenbogaard and Baron (1989). Internal wave breaking, like surface wave breaking, is most likely in the vicinity of the source of the waves and decreases as the wave energy is propagated to larger volumes of fluid where the wave mechanics become increasingly linear and wave shear rates increasingly non-intermittent, Phillips (1957), Munk (1981).

## Evidence of Saturated Internal Wave Motions Produced by Turbulence

Gargett et al. (1981) have published a composite spectrum of vertical shear in the upper ocean that shows a flat subrange out to wave lengths of about 10 meters, followed by a $k^{-1}$ subrange they attribute to saturated internal waves, and a bump they attribute to turbulence. Gibson (1986) has reinterpreted the $k^{-1}$ spectrum as subsaturated internal waves—decayed fossil vorticity turbulence waves that were produced by previous powerful turbulence events on scales up to 10 meter vertical amplitudes—and the bump as fossil vorticity turbulence. Thus, the smallest scale internal waves of the Gargett et al. (1981) composite spectra, and also the intermittency of their shear rate S (and $S^4$), are caused by turbulence, *not vice versa!* Gregg (1977) published a corresponding "schematic" spectrum of temperature gradients with a similar flat subrange at large scales, a $k^{-1}$ subrange attributed to saturated internal waves, and a bump at high wave numbers attributed to turbulence. This spectrum was also reinterpreted by Gibson (1986) in a similar way. The $k^{-1}$ subrange was far below the universal temperature gradient spectrum of the Gibson (1980) fossil turbulence theory, by a factor of ten, and the bump at the end was classified as fossil or active-fossil temperature turbulence.

Rodrigues-Sero and Hendershott (1997) have measured temperature gradient spectra at several stations downstream of tidal sills in the Gulf of California. Near the sills the spectra have the form of universal temperature spectra mixed by turbulence, but tens of kilometers downstream and days later the spectra take the Gregg (1977) form, except with ten times higher $k^{-1}$ subrange levels corresponding to the Gibson (1980, 1986) universal saturated internal wave spectrum rather than the lower Gregg schematic form. Clearly these saturated internal waves, and their universal saturated internal wave velocity and temperature spectra, are caused by turbulence. Their signatures appear to be decaying to the same subsaturated forms as those observed in the interior of the ocean by Gargett et al. (1981) and Gregg (1977), as shown in Figure 3.

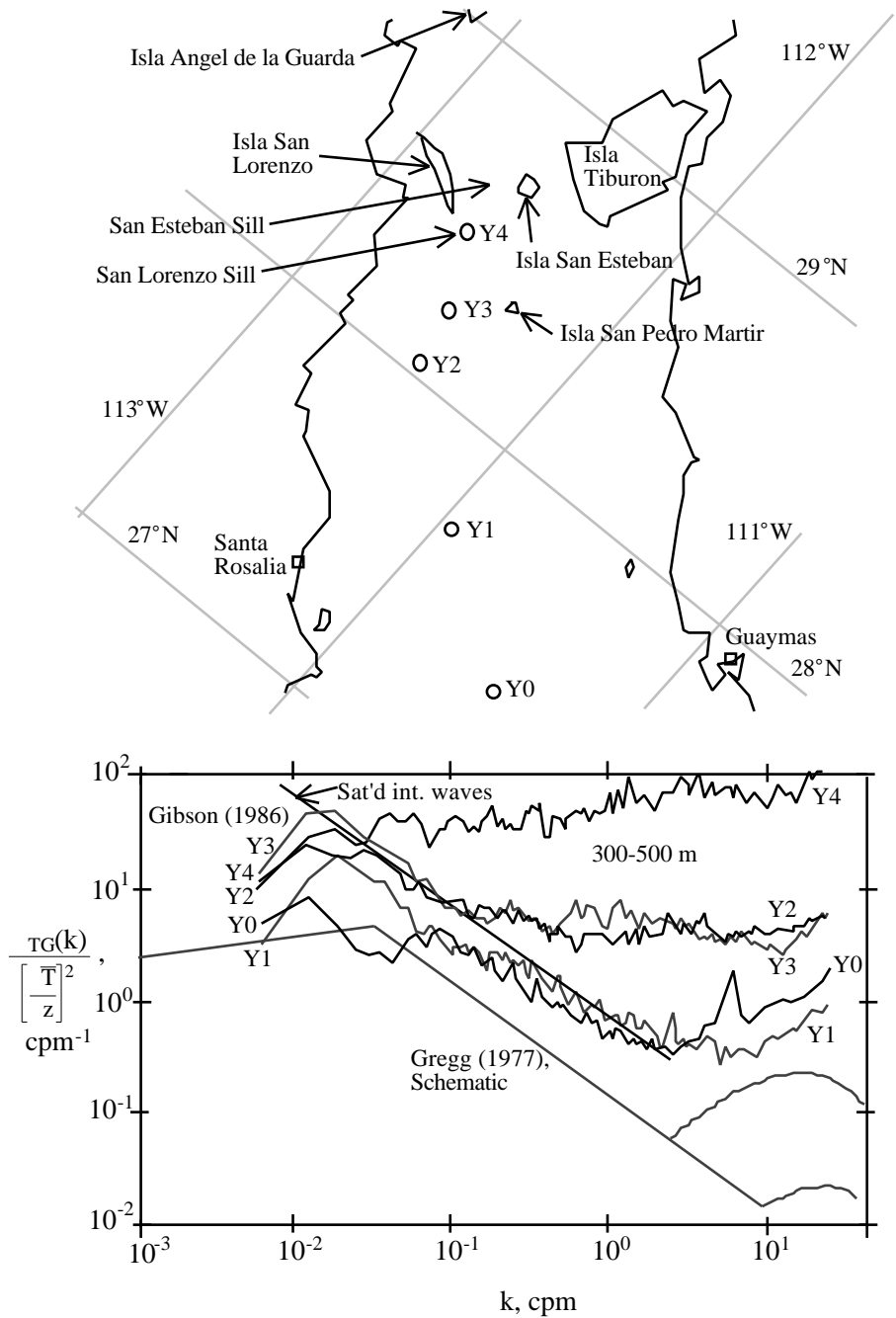

**Figure 3.** Temperature gradient spectra measured downstream of powerful turbulence events, with overturn scales up to 60 m, produced by flow over tidal sills. The spectra approach the universal form for saturated internal waves predicted by Gibson (1986), and are much higher than the schematic spectrum of Gregg (1977) which apparently represents subsaturated internal waves of the mid-Pacific thermocline. Measurements are from Rodrigues-Sero and Hendershott (1997).

Only if a persistent force maintains a tilted density interface long enough for significant shear to develop will nonlinearities and turbulence appear. When they do, the result is first the development of turbulence at the smallest scale possible for turbulence, the Kolmogorov scale, breaking the sheets of density gradient that produced the vorticity on tilting. Once the density concentration at the sheet has been broken, turbulence will cascade to larger scales and absorb all the kinetic energy that has accumulated on the tilted density surface while it has been tilted. If the tilting has been due to a passing internal wave, the patch of turbulence will be small because the maximum time period for kinetic energy accumulation is the internal wave period. If the tilting has gone on for time periods much longer than an internal wave period, as at a front, the concentration of density will be larger and the corresponding concentration of kinetic energy will be larger than the concentrations possible from large scale internal wave breaking. When the turbulence patch developed on a front reaches its maximum vertical size its kinetic energy is converted to the internal wave energy of fossil vorticity turbulence with persistence times much larger than the period of turbulence activity. Thus, the most extreme turbulence concentrations, and the most extreme values of small scale internal wave shears, will occur in association with fronts, and are caused by whatever causes the fronts. By this model, the small scale internal wave shear S, and the intermittency of $S^4$, are caused by turbulence, not the other way around. The small scale internal waves are fossil vorticity turbulence patches in various stages of decay from their initial saturated state and observed spectra, Gibson (1980, 1986).

Conclusions

Fronts in the ocean develop for a variety of reasons: including two-dimensional turbulence; a persistent source of density difference such as an island accumulating rain or a river emerging from a coastline; or a continental sized source of density difference like the polar caps. Two dimensional turbulence is defined as an eddy-like state of fluid motion where the inertial-vortex forces of the eddies in a [horizontal] plane are larger than any other forces that tend to damp them out, Gibson (1991cd), even though turbulence may be constrained at smaller scales in directions perpendicular to the plane. In the ocean interior far from islands or continents, two dimensional turbulence is the most likely source of fronts, and is therefore the most likely source of 3-D turbulence that causes the observed intermittency of small scale internal wave shear. The source of energy of two-dimensional turbulence is horizontal shear, such as boundaries between opposite currents in the ocean interior. The limitation of the size of two dimensional turbulence is presumably the Hopfinger scale $L_H = (\ /f^3)^{1/2}$ from Coriolis forces, where f is the Coriolis parameter, or the closely related Fernando scale $L_F = (q/f^3)^{1/2}$, where q is a buoyancy flux driving the turbulence. The smaller the Coriolis forces, the larger the horizontal scales of turbulence can become, the wider the range of scales in the turbulence cascade, and the larger the intermittency factors I according to [6]. This interpretation explains why the largest intermittency factors observed in the ocean, up to 7 or larger from Baker and Gibson (1987), are from equatorial data sets. Gregg et al. (1993) report intermittency factors for $S^4$ of over 6. If the Gregg (1989) correlation between    and $S^4$ is functionally correct (even though the estimated proportionality constant is not) then the extreme intermittency of $S^4$ observed can be attributed to the extreme intermittency of    resulting from a horizontal two-dimensional turbulence cascade.

The usual assumption that turbulence is caused by internal waves is questionable. Turbulence is caused by shear instability, and the largest shear instabilities in the ocean interior are those associated with fronts from two-dimensional turbulence with scales that can grow to tens of kilometers at midlatitudes and hundreds of kilometers at equatorial latitudes. The large intermittency factors of and $S^4$ observed in the ocean can both be attributed to the large range of scales of two dimensional turbulence following the Kolmogorov third hypothesis and [6]. Horizontal measurements of turbulence and internal wave statistics are needed to test this model of oceanic intermittency, preferably at various latitudes.

If the connection between turbulence intermittency and internal wave shear intermittency, and the correct cause-effect relationship, can be established, it may be possible to use the internal wave shear intermittency to infer turbulence intermittency factors that are needed in estimating turbulence transport processes in the ocean interior, Gibson (1987, 1990a, 1991bc, 1996abc).

## Acknowledgments

The author is grateful to Myrl Hendershott and Jim Brasseur for preprint versions of their very interesting papers, and to both of the referees for a number of helpful comments.